\documentclass[lettersize,journal]{IEEEtran}
\usepackage{amsmath,amsfonts}
\usepackage{algorithmic}
\usepackage{array}
\usepackage[caption=false,font=normalsize,labelfont=sf,textfont=sf]{subfig}
\usepackage{textcomp}
\usepackage{stfloats}
\usepackage{url}
\usepackage{verbatim}
\usepackage{graphicx}
\usepackage{booktabs} 
\usepackage[dvipsnames]{xcolor}
\usepackage{threeparttable}
\usepackage{multirow} 
\graphicspath{ {./images/} }
\def\BibTeX{{\rm B\kern-.05em{\sc i\kern-.025em b}\kern-.08em
    T\kern-.1667em\lower.7ex\hbox{E}\kern-.125emX}}
\usepackage{balance}
\usepackage{cite}
\usepackage{algorithm}
\usepackage{algorithmic}

\begin{document}
\bibliographystyle{ieeetr}
\title{Fuzzy Attention Neural Network to Tackle Discontinuity in Airway Segmentation}
\author{Yang Nan\textsuperscript{1},
        Javier Del Ser\textsuperscript{2,3},~\IEEEmembership{Senior Member,~IEEE,}
        Zeyu Tang\textsuperscript{1},
        Peng Tang\textsuperscript{4},\\
        Xiaodan Xing\textsuperscript{1},
        Yingying Fang\textsuperscript{1},
        Francisco Herrera\textsuperscript{5,6},~\IEEEmembership{Senior Member,~IEEE,}\\
        Witold Pedrycz\textsuperscript{7,8,9},~\IEEEmembership{Fellow,~IEEE},
        Simon Walsh\textsuperscript{1,10,*}, Guang Yang\textsuperscript{1,10,*},~\IEEEmembership{Senior Member,~IEEE}

\thanks{
\noindent *S. Walsh and G. Yang are the co-last senior authors\\
Send correspondences to: y.nan20@imperial.ac.uk, s.walsh@imperial.ac.uk, g.yang@imperial.ac.uk\\
        1. National Heart and Lung Institute, Imperial College London, London, UK\\
        2. Department of Communications Engineering, University of the Basque Country UPV/EHU, Bilbao, Spain\\
        3. TECNALIA, Basque Research and Technology Alliance (BRTA), Derio, Spain\\
        4. Department of Informatics, Technical University of Munich\\
        5. Department of Computer Sciences and Artificial Intelligence, Andalusian Research Institute in Data Science and Computational Intelligence (DaSCI) University of Granada, Granada, Spain\\
        6. Faculty of Computing and Information Technology, King Abdulaziz University, Jeddah, 21589, Saudi Arabia\\
        7. Department of Electrical and Computer Engineering, University of Alberta, Edmonton, Canada\\
        8. Department of Electrical and Computer Engineering, Faculty of Engineering, King Abgudulaziz University, Jeddah, Saudi Arabia\\
        9. The Systems Research Institute, Polish Academy of Sciences, Warsaw, Poland\\
        10. Royal Brompton Hospital, Sydney Street, London, UK}}

\markboth{IEEE Transactions on Neural Networks and Learning Systems}%
{How to Use the IEEEtran \LaTeX \ Templates}

\maketitle

\begin{abstract}
Airway segmentation is crucial for the examination, diagnosis, and prognosis of lung diseases, while its manual delineation is unduly burdensome. To alleviate this time-consuming and potentially subjective manual procedure, researchers have proposed methods to automatically segment airways from computerized tomography (CT) images. However, some small-sized airway branches (e.g., bronchus and terminal bronchioles) significantly aggravate the difficulty of automatic segmentation by machine learning models. In particular, the variance of voxel values and the severe data imbalance in airway branches make the computational module prone to discontinuous and false-negative predictions. especially for cohorts with different lung diseases. Attention mechanism has shown the capacity to segment complex structures, while fuzzy logic can reduce the uncertainty in feature representations. Therefore, the integration of deep attention networks and fuzzy theory, given by the fuzzy attention layer, should be an escalated solution for better generalization and robustness. This paper presents an efficient method for airway segmentation, comprising a novel fuzzy attention neural network and a comprehensive loss function to enhance the spatial continuity of airway segmentation. The deep fuzzy set is formulated by a set of voxels in the feature map and a learnable Gaussian membership function. Different from the existing attention mechanism, the proposed channel-specific fuzzy attention addresses the issue of heterogeneous features in different channels. Furthermore, a novel evaluation metric is proposed to assess both the continuity and completeness of airway structures. The efficiency, generalization and robustness  of the proposed method have been proved by training on normal lung disease while testing on datasets of lung cancer, COVID-19 and pulmonary fibrosis.
\end{abstract}

\begin{IEEEkeywords}
Airway segmentation, fuzzy neural networks, fuzzy attention, COVID-19, pulmonary fibrosis
\end{IEEEkeywords}

\section{Introduction}
\IEEEPARstart{P}{ulmonary} fibrosis occurs when the lung tissue becomes scarred and damaged, resulting in symptoms such as dyspnea, and is responsible for 1\% of all deaths in the UK. A challenge to managing patients with fibrotic lung disease is that currently there are no reliable means to predict progressive disease using baseline data.  The early identification of the progressive fibrotic lung disease would allow clinicians to initiate therapies to slow or prevent progression at the earliest opportunity, without the need to delay intervention until progression has been clinically observed. The lack of reliable discriminatory data at baseline is possibly the most urgent unmet challenge for effective management for patients with pulmonary fibrosis \cite{ley2014molecular,maher2017epithelial,maher2019biomarkers}.

\begin{figure}[tb]
\includegraphics[scale=1.3]{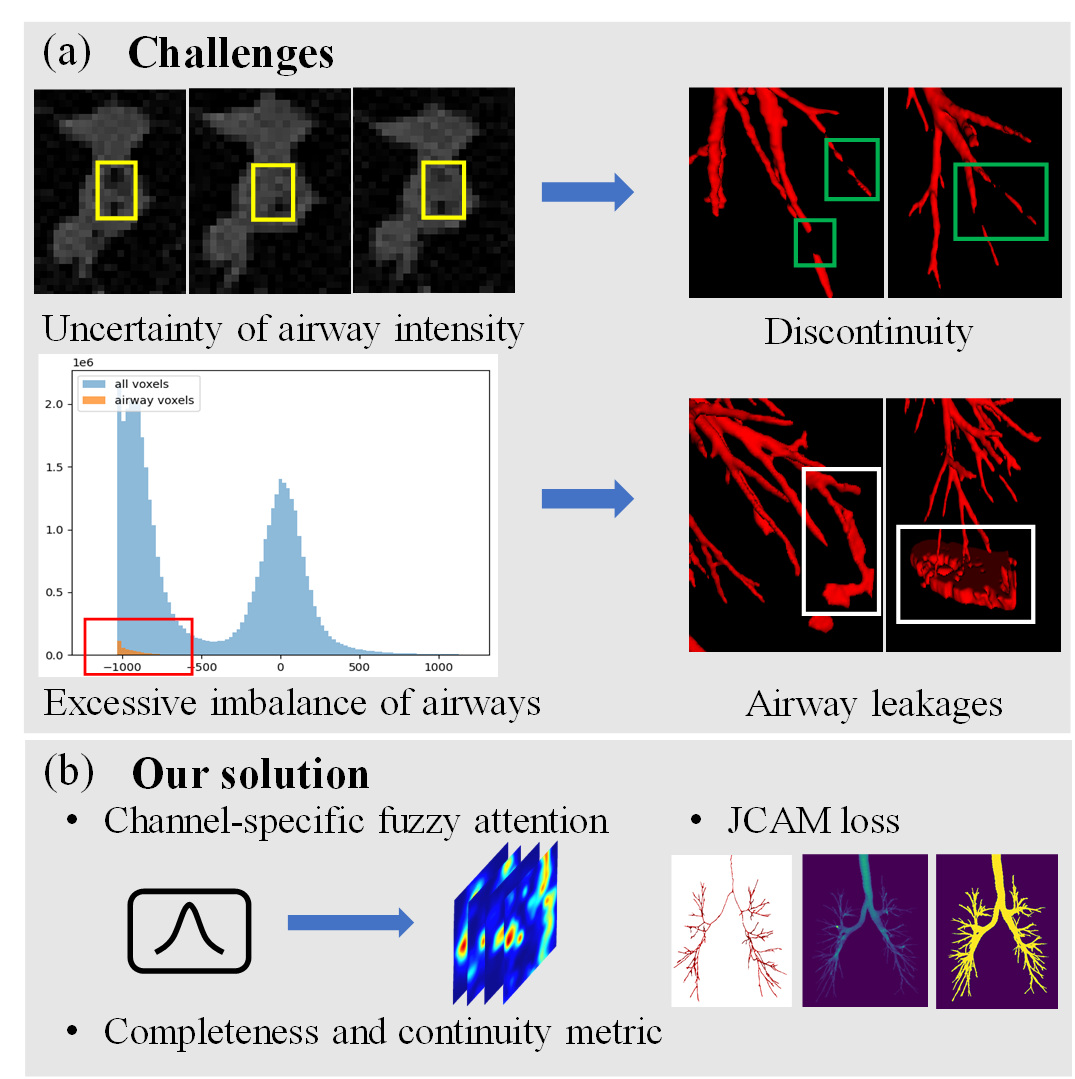}
\caption{Current challenges and our solution of airway segmentation. (a) The heterogeneity (yellow boxes) of the intensity of airway structures and the excessive data imbalance (red box), which leads to discontinuity predictions (green boxes) and leakages (white box); (b) This paper presents a fuzzy attention neural network to help the module to better learn the robustness feature, a JCAM (Jaccard Continuity and Accumulation Mapping) loss for optimization, and a novel metric that assesses the completeness and continuity of airway predictions.}
\vspace{-0.4cm}
\label{fig1}
\end{figure}

High resolution computed tomography (HRCT) of the chest is a key initial investigation in patients with suspected fibrotic lung disease and has been the focus of extensive biomarker research over the past 30 years. One HRCT biomarker which strongly predicts outcome in several distinct fibrotic lung subsets based on visual assessment is the severity of traction bronchiectasis \cite{walsh2012chronic,walsh2014connective,edey2011fibrotic}. Traction bronchiectasis is the abnormal dilatation of the tracheobronchial tree due to surrounding fibrosis. However, visual quantification of disease on computed tomography is liable to significant interobserver variability, and poor reproducibility and is relatively insensitive to subtle but clinically important changes over short follow up periods \cite{walsh2016interobserver}. This provides the rationale for developing objective computer-based methods for disease quantification on HRCT \cite{jacob2017mortality,humphries2017idiopathic, hansell2015ct}. Such methods can support many applications capable
of accurately quantifying lung diseases on HRCT: among them, we focus on airway segmentation, which refers to those computer-assisted methods that allow for an automated evaluation of the tracheobronchial tree. Unfortunately, airway segmentation remains elusive due to the high complexity of airway structures and inconsistent boundaries, especially for patients with fibrosis or COVID-19.

The feasibility of using conventional methods (e.g., morphology, intensity thresholding \cite{aykac2003segmentation}, region growing \cite{tschirren2005intrathoracic,5-kiraly2002three,6-graham2010robust}, fuzzy connectedness \cite{7-xu2015hybrid}) for airway segmentation was reported in a previous challenge EXACT-09 \cite{8-lo2012extraction}. However, these approaches were shown to suffer from weak robustness and coarse predictions due to the high complexity of bronchus patterns. Considering the airway segmentation task, there exist five major challenges (1) the imbalanced data distribution of foreground and background samples; (2) the homogeneity of voxel values between airway structures and normal lung regions; (3) the heterogeneity of voxel values in manual annotations and the incorrect labelling; (4) the weak reliability of evaluation metrics for module selection; (5) lack of generalization and robustness during the clinical pratice. Firstly, the huge imbalanced distribution of airway trees and other structures significantly aggravates the difficulty of training a data-driven based deep network. Most 3D neural networks are trained on patch datasets, while the criteria for extracting those patches are not valued. For instance, most studies simply extracted patches through a sliding window mechanism, while ignoring other sampling approaches. Secondly, the voxel values of normal lung regions and airway trees are similar, which requires the network to learn more spatial and structural information than voxel intensities. Thirdly, the inevitable incorrect manual annotations and the variances of voxel values in airway regions also require a strong robustness module. Moreover, the inappropriate metrics for module selection also hinder the saving of trainable weights. For example, most researchers employed intersection over union (IoU) or Dice score as the metric to save the best model. Unfortunately, the imbalanced voxel volume of the main trachea and small branches makes the overlapped metrics unreliable. Last but not least, most modules suffer from weak generalization, robustness, and require further fine-turning for different cohorts of different diseases.

Recently, deep neural networks have dominated the basic tasks and have become a common solution for biomedical image segmentation (e.g., the well known nnUNet \cite{28-isensee2021nnu} and VNet \cite{29-milletari2016v}). Based on the high modelling capacity of Convolutional Neural Networks (CNNs), researchers have achieved great efforts in airway segmentation \cite{9-charbonnier2017improving, 10-yun2019improvement,11-garcia2018automatic,12-qin2019airwaynet,13-cciccek20163d}. However, these methods still faced the challenge of low continuity and completeness of airway predictions. Among several deep learning techniques, the attention mechanism \cite{26-oktay2018attention} has shown a high capacity to segment complex structures. Meanwhile, studies have shown that by implementing fuzzy logic in deep neural networks, the uncertainty in feature representations can be reduced and significant improvements have been observed \cite{21-deng2016hierarchical,24-guan2019lip,shen2020hierarchical}. Therefore, we hypothesize the combination of deep attention mechanism and fuzzy theory should be able to tackle the challenges in accurate airway segmentation.

This paper presents an efficient scheme for accurate airway segmentation, including 1) smart patch sampling criteria, 2) a novel fuzzy attention neural network (FANN) for segmentation, and 3) a comprehensive loss function to enhance the continuity of airway predictions. Different from the previous methods that integrated fuzzy logic with deep learning through sequential learning paradigms (e.g., multi-module embedding or feature representation fusion), this paper incorporates fuzzy theory and attention mechanism as a fuzzy attention layer. The attention map in the fuzzy attention layer is formulated by a set of voxels in the feature map and a learnable Gaussian membership function, which is channel-specific, compared with the current channel-homogeneous attention mechanism. This fuzzy module can help the network better focus on the relevant feature representations while reducing their uncertainty to improve module's generalization and robustness. Meanwhile, an aggregative loss function that comprises airway continuity loss, accumulation mapping loss, and Dice loss is proposed. The airway continuity loss assesses the airway centerlines, whereas the accumulation mapping loss assesses the 3D prediction by estimating the error of mappings from different axes. Furthermore, a novel evaluation metric AF-score (airway f-score) is proposed to assess both the continuity and completeness of airway predictions.

Comprehensive experimental studies were conducted over an open-access dataset (90 cases) and our in-house datasets (40 cases), including normal scans, patients with mild anomalies, lung cancer, fibrosis, and COVID-19. The performance is assessed by calculating the completeness (dice coefficient score) and continuity (detected length and detected branches) of airway trees. In addition, the radius of each bronchus in the airway trees is estimated for a better understanding of the disease etiology. The preview of segmentation results is also demonstrated to provide a straight comparison.

The main contributions of this paper can be summarized as follows:
\begin{itemize}
    \item A smart patch sampling strategy is proposed to select appropriate patches for training deep networks.
    \item A novel channel-specific fuzzy attention layer by incorporating attention mechanisms and fuzzy logic for 3D segmentation. The deep fuzzy set is formulated by a set of voxels in the feature map and learnable Gaussian membership functions (MFs). To the best of our knowledge, this is the first attempt to adopt fuzzy logic in the attention layer.
    \item A comprehensive loss function including airway continuity loss and accumulation mapping loss is proposed to enhance the continuity and completeness of bronchus. 
    \item A novel evaluation metric CCF-score for module selection.
    \item Comprehensive evaluations of the proposed method and comparisons are given in terms of different lung diseases including pulmonary fibrosis cases, lung cancer and COVID-19.
\end{itemize}

\noindent The rest of this paper is organised as follows. The related works in airway segmentation and fuzzy logic are summarised in Section II. Details of the proposed method are illustrated in Section III. The experimental settings and results are described in Section IV.  Sections V and VI present the discussion and conclusion of this study.

\section{Related Works}
\subsection{Airway Segmentation in HRCT}
\noindent Segmentation of airway trees has always been a challenging task, due to the stated issues (in the introduction). E.g., the results in EXACT’09 demonstrated that no algorithm was capable to extract more than an average of 74\% of the total length of all branches in the reference. To obtain a better performance, Charbonnier et al. \cite{9-charbonnier2017improving} designed a 2D CNN to detect airway leakages for post-processing. Yun et al. \cite{10-yun2019improvement} presented a 2.5D CNN that takes three adjacent slices in each orthogonal direction (axial, sagittal, coronal) as inputs to increase the spatial information. In addition to 2D and 2.5D architectures, 3D CNNs  \cite{11-garcia2018automatic,12-qin2019airwaynet,13-cciccek20163d} became a popular solution due to the better consistency and integrity of their prediction.  

Some researchers regarded them as a class imbalance problem and proposed approaches to reduce interference of background regions. For instance, both Garcia-Uceda et al. \cite{14-garcia2021automatic} and  Qin et al. \cite{15-qin2020airwaynet} cropped the full-size volumes to a bounding box based on the pre-segmented lung region to remove the irrelevant background regions. Juarez et al. \cite{30-garcia2018automatic} applied weighted cross-entropy loss to balance the foreground and background samples. Zheng et al. \cite{16-zheng2021alleviating} trained the network on sampled patches in the first stage and selected hard-segment regions based on the false-negative predictions. Then, the network was fine-tuned in the second stage to further improve the performance.

Others aim to improve the prediction results by addressing the discontinuous issue and airway leakages. For example, Qin et al. \cite{12-qin2019airwaynet} proposed a 3D UNet with the voxel-connectivity awareness to segment airways. Meng et al. \cite{17-meng2017tracking} and Reynisson et al. \cite{18-reynisson2015airway} improved the continuity of predictions by learning the airway centrelines. Nadeem et al. \cite{19-nadeem2020ct} proposed 3D UNet with a freeze-and-grow propagation to reduce the airway leakages and improve the accuracy of detected branches. Zheng et al. \cite{16-zheng2021alleviating} analysed and explained the inaccurate segmentation issues by gradient erosion and dilation of adjacent voxels. Wang et al. \cite{31-wang2019tubular} presented a radial distance loss to enhance the completeness of tubular predictions.

However, even with these efforts, discontinuous predictions still exist, especially for those small airway branches. Moreover, most of these previous methods require multi-stage training protocol or complex post-processing procedures, which aggravates the computational and time costs in practical use. 

\vspace{-0.25cm}
\subsection{Fuzzy logic in Deep Semantic Segmentation}
\noindent In addition to the well-known fuzzy clustering approaches \cite{lei2019automatic, lei2018superpixel} for segmentation, there were many earlier attempts to combine deep neural networks and fuzzy logic through sequential learning protocol. For instance, using neural networks to extract low dimensional feature representations and applying fuzzified representations for classification \cite{20-zhou2014fuzzy}. Unfortunately, these multi-stage approaches cannot be trained as an end-to-end scheme and did not aggregate fuzzy theory into the training of deep neural networks.

To address this issue, some researchers integrated fuzzy logic with deep neural networks \cite{21-deng2016hierarchical} and proposed an end-to-end hierarchical scheme for deep fuzzy neural network. With these efforts, researchers have presented approaches to combining fuzzy logic with fully CNNs for segmentation \cite{22-price2019introducing,23-huang2021semantic,24-guan2019lip}. For instance, Price et al. \cite{22-price2019introducing} proposed a flexible and capable fuzzy layer to utilize the powerful aggregation of fuzzy integrals. Huang et al. \cite{23-huang2021semantic} proposed a fuzzy fully convolutional network for breast ultrasound image segmentation by transforming the data into the fuzzy domain. Guan et al. \cite{24-guan2019lip} proposed a fuzzy CNN for lip segmentation, introducing the fuzzy logic module to enhance the robustness of segmentation results. Ding et al. \cite{ding2021multimodal}. presented fuzzy-enabled multi-scale learning to segment brain tumours from T1 and T2 scans simultaneously. However, most existing methods adopted fuzzy integral within a residual block for information fusion, which cannot address the main challenges of airway segmentation (shown in the experiment section). Besides, only the ‘AND’ aggregation operator was applied to membership functions in most previous studies without exploring other operators.
\vspace{-0.25cm}
\subsection{Evaluation Metrics of Airway Predictions}
\noindent Given the semantic prediction $X$ and ground truth annotation $Y$, the main evaluation metrics for airway segmentation include overlapped metrics, detected branch ratio, detected length ratio and airway leakage ratio, which are given as follows:
\begin{enumerate}
    \item \textbf{Overlapped metrics} measure the percentage overlap between the prediction and ground truth, including dice coefficient score and Jaccard index score (or intersection over union score), which are represented as:\\
    \begin{equation}
    \mathrm{Jaccard}(X,Y)=\frac{XY}{X+Y-XY},
    \end{equation}
    \begin{equation}
    \mathrm{Dice}(X,Y)=  \frac{2XY}{X+Y},
    \end{equation}
    where $XY$ are calculated by dot product of $X$ and $Y$.
    \item \textbf{Detected branch ratio (DBR)} measures the proportion of detected branches with respect to the ground truth annotations\\
    \begin{equation}
        \mathrm{DBR}=\frac{N_x}{N_y} \times 100\%,
    \end{equation}
    where \(N_X\)  is the number of branches that have been correctly recognized, \(N_Y\) is the number of branches in the ground truth annotation. In this study, branches with the intersection over union (IoU) score greater than 0.8 are referred to be correctly identified.
    \item \textbf{Detected length ratio (DLR)} measures the proportion of detected branch length with respect to that of the ground truth annotations\\
    \begin{equation}
        \mathrm{DLR}=\frac{L_x}{L_y}\times100\%,
    \end{equation}
    where \(L_X\) is the total length of branches that have been correctly recognized, \(L_Y\) is the total length of branches in the ground truth annotation.
    \item \textbf{Airway leakage ratio (ALR)} refers to the proportion of total false positive volumes with respect to the ground truth annotations\\
    \begin{equation}
        \mathrm{ALR}=\frac{V_x}{V_y}\times100\%,
    \end{equation}
    where \(V_X\) is the volume of false-positive predictions, \(V_Y\) is the volume of ground truth annotation.
\end{enumerate}
Though the airway segmentation can be evaluated through these metrics, none of them can be used independently for assessment. 

\section{Methodology}
\noindent This section illustrates details of the proposed scheme for airway segmentation, comprising the overview of FANN, the smart patch sampling, a fuzzy attention layer, a JCAM (Jaccard continuity and accumulation mapping) loss and a CCF-score (Continuity and Completeness F-score).
\subsection{Overview}
\begin{figure}[h]
\vspace{-0.3cm} 
\includegraphics[scale=0.65]{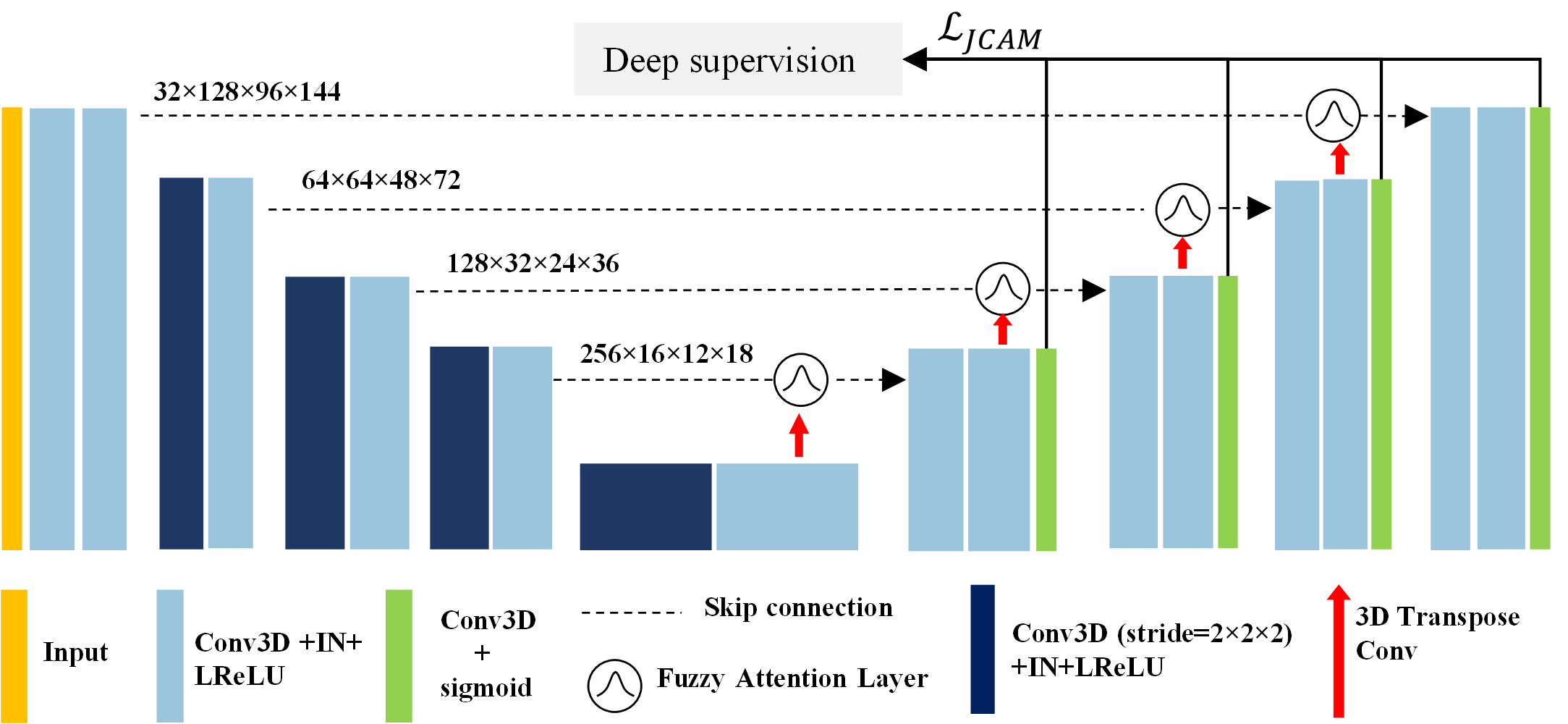}      
\centering
\caption{Fuzzy attention Neural Network. The fuzzy attention layer is adopted within the skip connection between the encoders and decoders. Each decoder block produces a binary prediction for deep supervision. Sizes of feature maps are presented for a better understanding of the architecture of FANN.}
\label{fig2}
\end{figure}

\noindent Given an input raw 3D volume \(V\in{\mathbb{R}^{W\times{H}\times{Z}}}\), a smart patch sampling strategy is first presented to extract appropriate patches \(V_P\in{\mathbb{R}^{w\times{h}\times{c}}}\) for training and validation. Then, a fuzzy attention neural network (FANN) is proposed to segment the airway trees with the supervision of a comprehensive loss function (including dice loss, airway continuity loss and accumulation mapping loss) on \(V_P\). During the training, the AF-score is developed to assess the model performance and save the best values of the trainable network parameters. 

The proposed FAAN is built based on the 3D U-Net by adding a novel fuzzy attention layer, deep supervision, and airway continuity and accumulation mapping (ACAM) loss. The size of the volume patches \(V_P\) is set as the same ratio as the median shape of the ground truth annotations in the training data. FANN includes 3D convolution layers, instance normalization (IN), LeakyReLu (LReLU), 3D transpose convolution layers, fuzzy attention layers and sigmoid activation layers, as shown in Fig. 2. It is of note that the proposed network has multiple outputs, including the main output and three low-level outputs. The final prediction is given from the main output, while the low-level outputs are collected for deep supervision through the auxiliary losses (losses are calculated as the same way as that of the main output with different weights). When calculating these auxiliary losses, the prediction is upsampled to align with the size of ground truth masks.

\vspace{-0.3cm} 
\subsection{Smart Patch Sampling (SPS)}
\noindent  Deep learning is a data-driven method that highly relies on a large amount of data, thus a balanced class distribution is crucial for training a neural network. The huge imbalance of foreground and background samples are prone to gradient erosion issue \cite{16-zheng2021alleviating}. Due to the large volume array in HRCT and GPU memory limitations, segmentation modules in HRCT are mainly based on patch training strategies. However, most studies simply extracted patches through sliding windows, which cannot completely address the class imbalanced issue. Although some researchers alleviated class imbalance by extracting hard samples \cite{16-zheng2021alleviating, add1-xie2020relational}, these methods required a further training process to collect hard samples. In this study, we present smart patch sampling (SPS) criteria for patch extraction that do not require further training/fine-tuning operations. Given the dataset that comprises 3D images and their corresponding annotations, the workflow of SPS can be summarized as:
\begin{enumerate}
    \item 	The average size S (\(z\times{y}\times{x}\)) 	of the 3D minimum bounding volume of ground truth annotations is first calculated. The patch size is set as the same scale as S. 
    \item   The centreline  of manual annotation is extracted by skeletonization \cite{25-lee1994building}.
    \item   Overlapped sliding windows are adopted to extract image patches, mask patches, and centreline patches. 
    \item   4)	Patches with a centerline ratio (the proportion of the centerline voxels in a patch in terms of overall centerline voxels) larger than 15\% or a voxel volume ratio (the proportion of the ground truth voxels in a patch in terms of overall ground truth voxels) larger than 10\% are kept.
\end{enumerate}

\vspace{-0.3cm} 
\subsection{Fuzzy Attention Layer}
\noindent One challenge towards obtaining a well-trained module for airway segmentation is the uncertainty of annotations and voxel values within the airway region. There have been efforts to make the network focus more on the relevant regions. For instance, the Attention U-Net \cite{26-oktay2018attention} proposes an attention gate to improve the accuracy by suppressing feature activations in irrelevant regions. However, we deem that the sigmoid activation may not be the best solution to organize the attention gate. 

In addition to the "gradient vanishing" problem, one major concern with the sigmoid activation function is its sharpness, by which only a small interval can obtain the output value ranges between 0 and 1. Therefore, it is difficult to find a robust "boundary" in sigmoid activation that distinguishes whether the feature is relevant or not. Another issue is the monotonicity: similar to the raw intensity distribution of airway structures that has both negative and positive variations, the feature representation of airway regions should also have two side variations. However, to align with the sigmoid activation function, the \(1\times1\) convolution layers must learn to shift the values of all the features of interest to a single side to obtain a positive response. In particular, there must exist a certain "threshold" in the feature representation reconstitution (usually accomplished by the \(1\times1\) convolutional layers in the attention layer) to determine whether the region is important or not. Moreover, non-channel specifics of the current attention map assign the same "attention" coefficient to all the feature points along the channel-wise. Specifically, given a feature representation \(F\in{\mathbb{R}^{C\times{W}\times{H}\times{D}}}\), the existing attention map is formulated as \(\alpha\in{\mathbb{R}^{W\times{H}\times{D}}}\), while all the feature representations along the channel wise C shares the same ‘importance’. This mechanism is unreliable since the feature representations in different channels are extracted by different convolution kernels; therefore, we advocate the attention map to be channel-specific. 

\begin{figure}[t]
\includegraphics[scale=0.9]{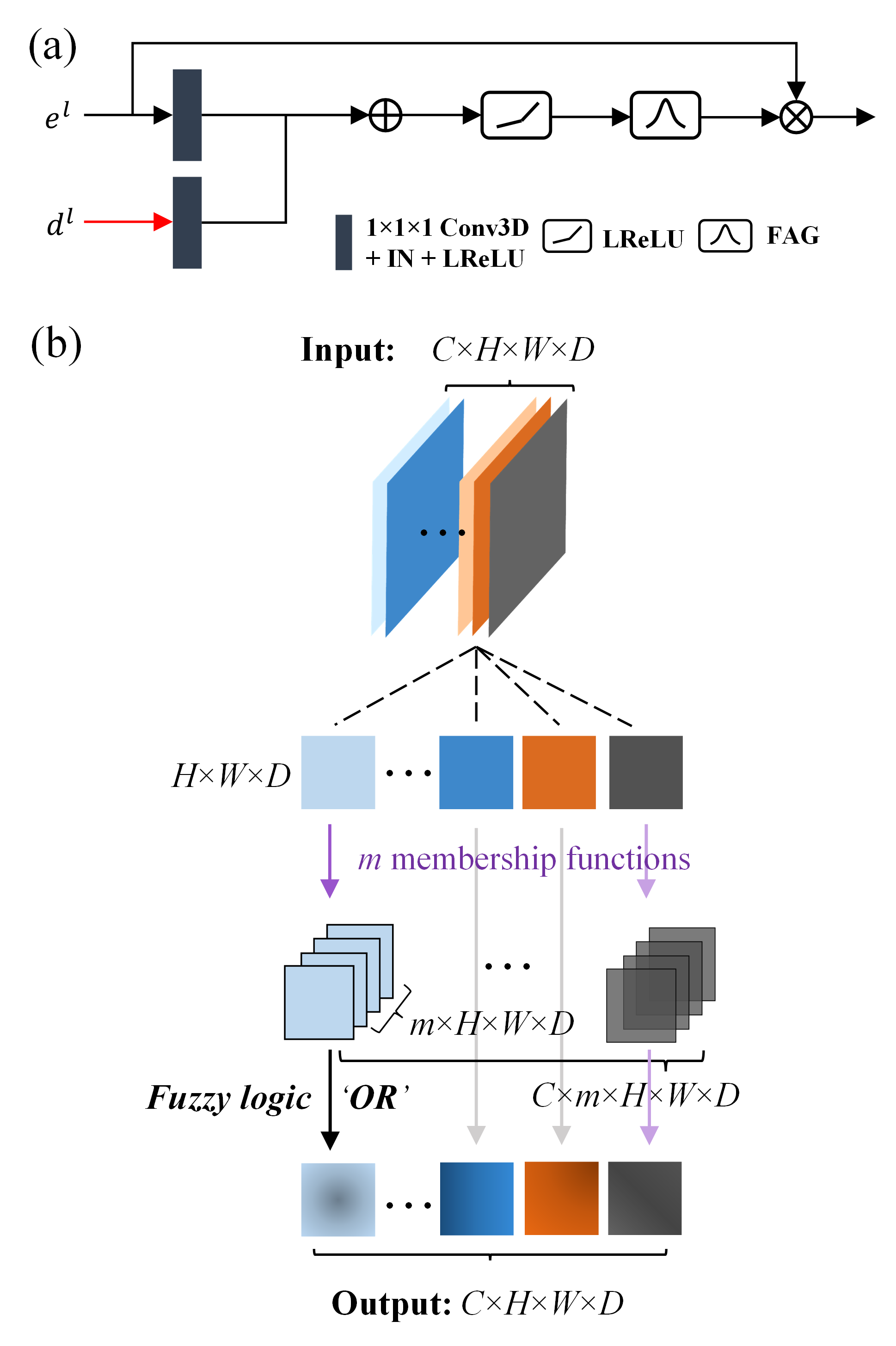}      
\centering
\caption{Details of (a) fuzzy attention layer and (b) fuzzy attention gate (FAG). The fuzzy attention layer takes both feature representations from the \textit{l-th} encoder layer \(e^l\) and decoder layer \(d^l\) as inputs and outputs an attention map that has the same dimension as the inputs.The learnable Gaussian membership function is adopted.}
\label{fig3}
\vspace{-0.4cm} 
\end{figure}

Different from the sigmoid activation function which is fixed and sharped, we believe the Gaussian function is more effective to formulate the attention gate, due to its symmetry and flexibility (the mean and variance of the Gaussian function can be set as learnable parameters). Studies have shown that both fuzzy logic and neural networks are efficient for data representation  \cite{21-deng2016hierarchical}. In general, the neural network aims to reduce noises in the original data to extract useful feature representations, whereas fuzzy logic is used to obtain fuzzy representations to reduce the uncertainty in original data. Therefore, we combine fuzzy logic with the attention mechanism using trainable Gaussian membership functions to better help the segmentation network focus on the relevant region while reducing the uncertainty and variations of the data representations.

The proposed fuzzy attention layer is adopted within the skip connection, taking both feature representations from the \textit{l-th} encoder layer space \(e^l\) and decoder layer space \(d^l\) as inputs (shown in Fig. 3(a)). These two input feature vectors are first processed by a \(1\times1\times1\) 3D convolution layer, an instance normalization, and a LeakyReLU for feature reconstitution. Then, a voxel-wise adding operation is adopted to fuse the information, followed by a LeakyReLU. Next, the feature representations are fed into the FAG to generate a voxel-wise attention map, shown in Fig. 3(b). Assume \(X\) (with a shape of \(C\times{H}\times{W}\times{D}\) regardless of batch size) as the input of the fuzzy attention gate. Due to the smoothness and concise notation of Gaussian membership functions, learnable Gaussian MFs are proposed to specify the deep fuzzy sets. Each feature map (with the height \(H\), width \(W\) and depth \(D\)) is filtered by \(m\) Gaussian membership functions with the trainable centre \(\mu_{i,j}\) and spread \(\sigma_{i,j}\)
\begin{equation}
    f_{i,j}(X,\mu,\sigma)=e^{\frac{-(X_j-\mu_{i,j})^2}{2\sigma_{i,j}^2}},
\end{equation}
where $i=1,…,m,j=1,…,C$. Different from most previous studies that applied the operators ‘AND’ to obtain fuzzy feature representations, our goal is to use the membership function to learn the ‘importance’ of target feature representations. Therefore, we believe that the information can be better preserved by applying the aggregation operator ‘OR’ while suppressing irrelevant features. Having two fuzzy sets \(\Tilde{A}\) and \(\Tilde{B}\), the operator ‘OR’ is described as
\begin{equation}
    f_{\Tilde{A}\cup\Tilde{B}}(y)=f_{\Tilde{A}}(y)\vee f_{\Tilde{B}}(y),~ \forall{y\in{U}},
\end{equation}
where U is the universe of information and y is the element of the universe. To make the operator ‘OR’ derivative, we modified it as 
\begin{equation}
    f_{\Tilde{A}\cup\Tilde{B}}(y)=\max(f_{\Tilde{A}}(y), f_{\Tilde{B}}(y)),
\end{equation}
Therefore, the fuzzy degree \(f_j(X,\mu,\sigma)\in{\Theta^{H\times{W}\times{D}}}, \Theta\in{[0,1]}\)
of \(j-th\) channel can be obtained based on Eq.(6) and Eq.(8) as
\begin{equation}
\begin{aligned}
     f_{j}(X,\mu,\sigma)
     &=\bigvee_{i=1}^{m} {e^{\frac{-(X_j-\mu_{i,j})^2}{2\sigma_{i,j}^2}}}=\max(e^{\frac{-(X_j-\mu_{i,j})^2}{2\sigma_{i,j}^2}})
\end{aligned}
\end{equation}
where the large V indicates the union operation.

\noindent Eventually, the output tensor of the proposed FAG has the same shape as the input tensors, which can provide a voxel-wise attention map. The pseudo-code of the fuzzy attention layer is shown in Algorithm 1.

\begin{algorithm}
	\renewcommand{\algorithmicrequire}{\textbf{Input:}}
	\renewcommand{\algorithmicensure}{\textbf{Output:}}
	\caption{Pseudo code of fuzzy attention layer}
	\label{alg1}
	\begin{algorithmic}[1]
	    \REQUIRE feature representations \(e^l, d^l \in{\mathbb{R}^{C\times{W}\times{H}\times{D}}}\),from l-th encoder and decoder, , weights \(w_{e}^{(l)},w_{d}^{(l)}\) and bias \(b_{e}^{(l)},b_{d}^{(l)}\) connecting to l-th encoder and decoder
		\ENSURE fused feature representation \(y \in{\mathbb{R}^{C\times{W}\times{H}\times{D}}}\)
		\STATE randomly initialise parameters  $\mu\in \mathbb{R}^{m\times C}$ and  $\sigma\in \mathbb{R}^{m\times C}$ of membership functions $f(X, \mu,\sigma)$.
		\STATE  \textbf{compute} the input $X\in \mathbb{R}^{C\times H\times W\times D}$ of FAG, \(X= \mathrm{LReLU}[\mathrm{LReLU}(\mathrm{IN}(w_e^{(l)}e^l+b_e^{(l)}))\) +\\ \(\mathrm{LReLU}(\mathrm{IN}(w_d^{(l)}d^l+b_d^{(l)}))]\)
		\FOR{i in C}   
		\STATE \textbf{compute} the fuzzy membership degrees $f_{j}(X,\mu,\sigma)$ through m membership functions \\
		\STATE \textbf{compute} degrees $f_{j}(X,\mu,\sigma)$ of j-th channel by adopting aggregation operator "OR"  to $f_{j}$ \\
		\STATE \textbf{compute} output of $j$-th channel \(y_j=e^{(l)}f_j(X, \mu,\sigma)\)
		\ENDFOR
	\end{algorithmic}  
\end{algorithm}

\vspace{-0.3cm} 
\subsection{Jaccard Continuity and Accumulation Mapping Loss}
\noindent Studies have shown no qualms about applying strategies such as cascade modules or cropping operations to better segment small organs or structures. By implementing these strategies, the inter-class imbalance issue can be well alleviated. However, another issue that the model needs to address is the intra-class imbalance, which is caused by different volumes of different branches (trachea, secondary/tertiary bronchus). The trachea and the secondary bronchus account for the majority of the total airway volume, leading to a discontinuous prediction for the small bronchus. 

\begin{figure}[hb]
\vspace{-0.3cm} 
\includegraphics[scale=0.8]{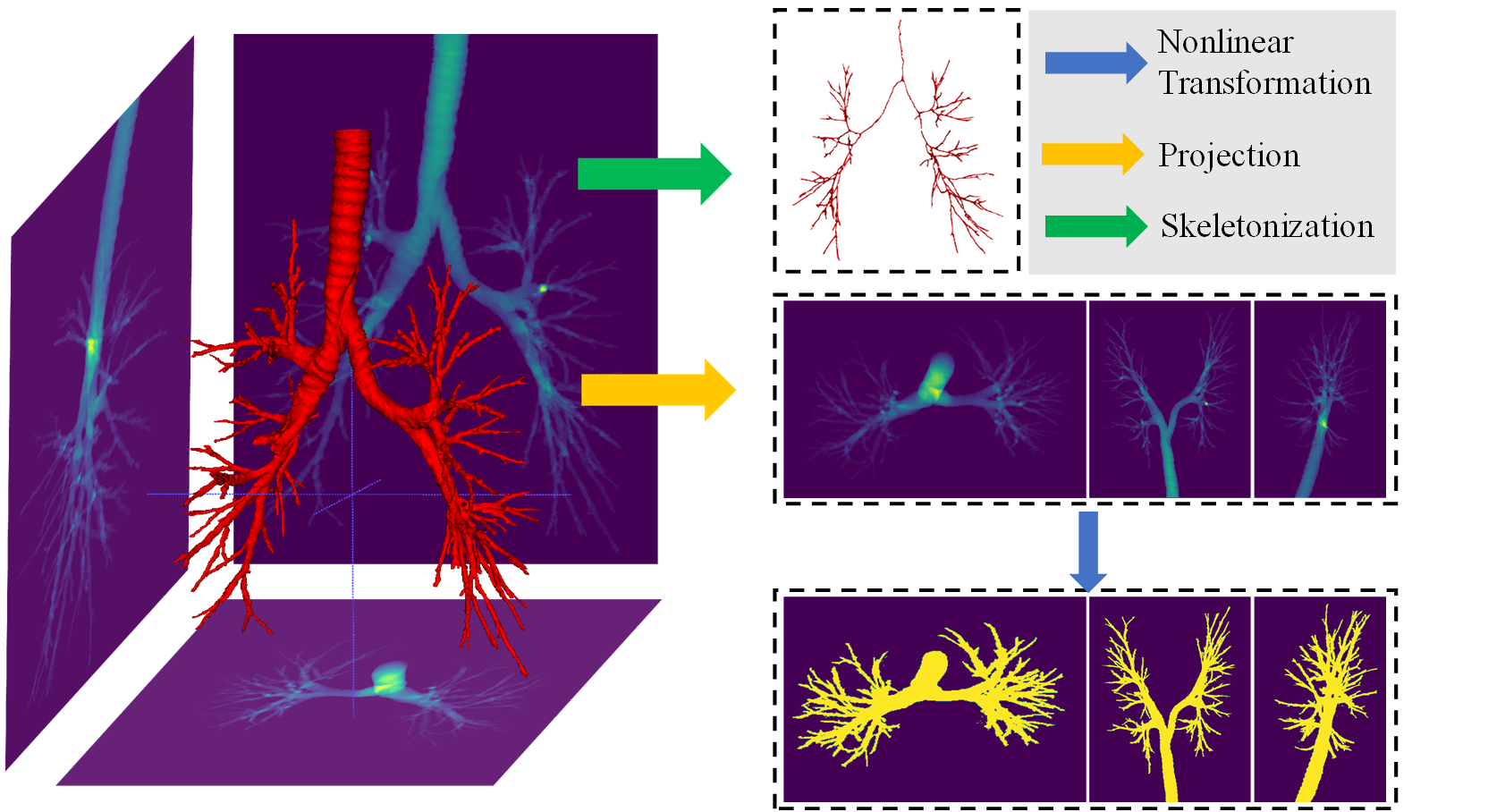}      
\centering
\caption{Skeletonization and accumulation maps (acquired from the projection of coronal, sagittal and transverse planes, and nonlinear transformation) of the airway structure.}
\vspace{-0.3cm} 
\label{fig4}
\end{figure}

To bridge this gap, we propose a Jaccard continuity and accumulation mapping loss \(L_{JCAM}\) to force the network to pay more attention to the continuity of airway predictions. The \(L_{JCAM}\)  assesses the error of different projections (through coronal, sagittal, and axial planes) and centrelines between the prediction and ground truth. Given the prediction array $X$, the corresponding ground truth annotation $Y$, the centrelines \(Y_{CL}\) of the ground truth masks are extracted by the skeletonization \cite{25-lee1994building}. With the ground truth centrelines \(Y_{CL}\), the continuity can be well evaluated by calculating the ratio of correct-predicted centrelines and ground truth centrelines
\begin{equation}
\begin{aligned}
     L_C
     &=1-C=1-\frac{\sum{X \cdot Y_{CL}}}{\sum{Y_{CL}}}
\end{aligned}
\end{equation}
without considering the airway branch size. In addition to the continuity, we also present two types of accumulation maps, including linear accumulation maps (LAMs) and nonlinear accumulation maps (nLAMs), as shown in Fig. 4. LAMs are calculated by summing the volume array along three different axes, while nLAMs are acquired by performing a nonlinear transformation on LAMs. By integrating LAM into the optimization function, the veracity of the 3D prediction can be better assessed since each incorrect voxel leads to incorrect variations in different projection maps. Furthermore, to reduce the intra-class imbalance, a nonlinear transformation (tanh activation function) is introduced to LAMs to acquire nLAMs.  Although nLAM  (ranges from [0, 1]) discards the spatial information within LAM, it can better gauge the continuity and veracity between the predictions and ground truth masks.

Denote \(A\oplus{\tau}\)  as the summation operation that sums the array \(A\in{\mathbb{R}^{N\times{C}\times{W}\times{H}\times{D}}}\) along \(\tau{th}\) channel. The LAM and nLAM are calculated by
\begin{equation}
    L_{LAM}=\sum_{\tau\in[W,H,D]}l_1(X\oplus{\tau}, Y\oplus{\tau}),
\end{equation}
\begin{equation}
    L_{nLAM}=\sum_{\tau\in[W,H,D]}L_{\mathcal{J}}(\tanh({X\oplus{\tau}),\tanh(Y\oplus{\tau}))},
\end{equation}
where \(L_\mathcal{J}\) is the Jaccard loss which is defined as
\begin{equation}
\begin{aligned}
     L_\mathcal{J}
     &=1-\mathcal{J}=1-\frac{XY+\varepsilon}{X+Y-XY+\varepsilon}
\end{aligned}
\end{equation}
and \(\varepsilon\) is the smoothing factor \(\varepsilon \textgreater 0\).

Overall, the \(L_{JCAM}\) can be summarised as 
\begin{equation}
\begin{aligned}
\begin{split}
    &L_{JCAM}=\alpha{L_\mathcal{J}}(X,Y)+\beta{L_C}(X,Y_{CL})+\\
    &\varphi{L_{CE}}(X,Y)+\gamma{L_{LAM}}(X,Y)+\delta{L_{nLAM}}(X,Y)
\end{split}
\end{aligned}
\end{equation}
where \(\alpha, \beta,\gamma,\varphi, \delta\) are the weights of Jaccard loss \(L_\mathcal{J}\), continuity loss \(L_C\), cross-entropy loss \(L_{CE}\), linear and nonlinear accumulation mapping loss \(L_{LAM}\), \(L_{nLAM}\), repectively.
\vspace{-0.3cm} 
\subsection{Continuity and Completeness F-score (CCF-score)}
In addition to developing novel layers or operations, module selection also plays a crucial role in machine learning. In airway segmentation, it is unreliable to find the best model by selecting the highest regional overlapping metrics, due to (1) the intra-class imbalance of the airway structures and (2) the imperfection of manual annotations. A high Dice or IoU score only reflects the overall overlapping ratio of the predictions voxels, which cannot assess the continuity and completeness. Thus, there is an urgent requirement to develop an effective metric to evaluate the continuity and completeness of tree structures (e.g., airways, blood vessels, neurons, etc.).

Here we propose a continuity and completeness F-score (CCF-score) for the aforementioned purpose
\begin{equation}
    \mathrm{CCF}_s=(1+\omega^2)\times \frac{{J}\times{C}}{\omega^2\times{J}+C}
\end{equation}
where \(\omega \in [0,1]\) is the preference parameter, \(J\) and \(C\) are the Jaccard index and Continuity index as defined in Eq.(10) and Eq. (13), respectively. \(\omega\) can be set larger (smaller) than 1 when \(C(J)\) is more important. With this novel F-measure based metric, the module that focuses on both continuity and completeness can be saved.

\section{Experiments and Results}
\noindent This section details all the settings of experiments including datasets, implementation details, evaluation metrics and results.  
\vspace{-0.3cm} 
\subsection{Datasets and Training Strategies}
\noindent In this study, we trained our airway segmentation and comparison models using 90 clinical thoracic CT scans from the open-access BAS dataset. In particular, the in-house datasets collected from patients with fibrotic lung disease and COVID-19 are used for testing only. Details of all the data are shown in Table I.

\begin{table*}[htb]
\centering
\caption{Details of datasets in this study}
\label{tab:1}  
\begin{threeparttable}
\begin{tabular}{lllll}
   \toprule
   Dataset & Composition & Volume size & Slice thickness & Disease \\
   \midrule
   BAS & 90 cases* & $s$×512×512, $s\in [157, 764]$  & 0.45mm-1.80mm & Healthy volunteers or patients with pulmonary disease\\
   COVID-19 & 25 cases (test only) & $s$×512×512, $s\in [512, 1145]$  & 0.40mm-0.65mm & Patients with COVID-19\\
   Fibrosis & 25 cases (test only) & $s$×512×512, $s\in [516, 945]$  & 0.40mm-0.70mm & Patients with fibrotic lung disease\\
  
   \bottomrule
\end{tabular}
\begin{tablenotes}
\footnotesize
\item * 90 cases are split to 54,18 and 18 for training, validation and testing, respectively.
\end{tablenotes}
\end{threeparttable}
\vspace{-1.5em}
\end{table*}

\noindent \textbf{Binary airway segmentation (BAS) dataset: }Overall, there are 90 cases in total for BASD with 20 cases from EXACT’09 and 70 cases from LIDC. (1) EXACT’09 \cite{8-lo2012extraction} is a dataset that consists of 20 cases for training (with the corresponding annotations) and 20 cases for the test (without annotation publicly available), with the scanned patients ranging from normal conditions to severe pulmonary lung diseases. (2) LIDC dataset: The Lung Image Database Consortium image collection (LIDC-IDRI) includes 1018 cases of diagnostic and lung cancer screening thoracic CT scans \cite{27-armato2011lung}. Among the 1018 cases, 70 cases whose slice thickness $\leq 0.625 mm$ were randomly selected and annotated by experts \cite{15-qin2020airwaynet, 16-zheng2021alleviating}.\\
\textbf{Our in-house dataset: }(1) COVID-19 dataset: this subset includes 25 HRCT for patients with COVID-19, from Wuhan Renmin hospital. (2) Fibrosis dataset: this subset contains 25 cases from patients with fibrotic lung disease, collected from OSIC dataset\footnote{https://www.osicild.org/}. The annotations of these cases were given by experts from Royal Brompton Hospital.\\
\textbf{Training Strategies:} Parameters of the proposed 3D-UNet are initialised with He-normal initialisation. Randomised rotation (rotation degree ranges from -10 to 10), and randomised flip (up, down, left, right) were implemented to augment the dataset during the training. All modules were trained on an NVIDIA RTX 3090 GPU for 200 epochs, with an initial learning rate of \(1e^{-3}\) and a decay of 0.5 at the 20th, 50th, 80th. 110th and 150th epoch. For equal contribution of \(L_\mathcal{J}, L_C, L_{CE}\) the value of the hyper-parameters $\alpha$, $\beta$, $\gamma$ in  $L_{JCAM}$ was set to 1, and that of $\varphi$, $\delta$ was set to 0.3 to constraint the accumulation mapping loss (since it is calculated through three dimension). The $\omega$ in CCF-score was set to 0.9 to prevent excessive leakages.

\subsection{Experimental Settings and Evaluation Metrics}
\noindent This section describes the experimental settings of airway segmentation. We randomly divided all the 90 cases from the BAS dataset into 72 for training and validation, and 18 for the test. All the modules in ablation studies and comparison models were trained from scratch based on the same patch-set given from SPS, using the same optimizer and training strategy. 

To provide a fair comparison, all the models were applied the same postprocessing (extracting the largest connected component to remove noises) and inferencing criteria (non-overlapping sliding window). The significance of the results between the proposed method and comparison method was assessed by Wilcoxon signed-rank test, with the p-value<0.05 indicating significant better(worse) results. 

\noindent \textit{\textbf{Ablation studies}}. To better evaluate the effectiveness of the proposed submodules, we conducted ablation studies for airway segmentation including 3D U-Net (BL), 3D U-Net with fuzzy attention layer (BL+Fuzzy), 3D U-Net with fuzzy attention layer and JCAM loss (BL+Fuzzy+JCAM), 3D U-Net with fuzzy attention layer, JCAM loss and CCF-score (BL+Fuzzy+JCAM+$\rm CCF_{s}$) and other combinations (BL+JCAM, BL+Fuzzy+$\rm CCF_{s}$). It is of note that the baseline 3D UNet has the same structure as the FAAN without a fuzzy attention layer (shown in Fig. 2) for the sake of fair comparison.\\
\textit{\textbf{Comparisons}}. To evaluate the effectiveness of the proposed method, we compare our method with existing approaches including nnUNet \cite{28-isensee2021nnu}, VNet \cite{29-milletari2016v}, Attention UNet\cite{26-oktay2018attention}, and methods proposed by Juarez et al. \cite{30-garcia2018automatic}, Zheng et al. \cite{16-zheng2021alleviating} and Wang et al. \cite{31-wang2019tubular}. All the modules were trained on the same patch dataset, acquired based on SPS criteria. Besides, all the comparisons were performed on open BAS and our in-house datasets to assess the robustness and performance, respectively.\\
\textit{\textbf{Evaluation metrics}}. In this study, we applied 5 metrics to evaluate the performance of airway segmentation modules. In addition to the three metrics (IoU, detected length ratio, detected branch ratio) in section 2.2, the precision and airway missing ratio (AMR) were also evaluated:
\begin{equation}
    \mathrm{Precision}=\frac{TP}{TP+FP}
\end{equation}
\begin{equation}
    \mathrm{AMR}=\frac{FN}{V_Y}\times100\%
\end{equation}
where $TP$, $FP$, and $FN$ are the number of true positive volumes, false positive volumes and false negative volumes.
\vspace{-0.3cm} 
\subsection{Experimental Results}
\noindent Results of all the experiments are calculated on the test cases shown as mean ± standard deviation in this section.\\
\textit{\textbf{Ablation studies}}. Results of the ablation experiments are presented in Table II, illustrating the mean and standard deviation of the different evaluation metrics. 
\begin{table*}[htb]
\centering
\caption{Ablation studies of the proposed method on bas dataset}
\label{tab:2}  
\begin{threeparttable}
\begin{tabular}{lllllll}
   \toprule
   Model & IoU & Precision & DLR(\%) & DBR(\%) & AMR(\%) & $\rm CCF_{s}$ \\
   \midrule
   BL & \textbf{0.8749±0.0385} & 0.9190±0.0302 & 87.81±8.74 & 81.74±11.15 & 5.11±4.10 & 0.8763±0.0514 \\
   BL+Fuzzy & 0.8731±0.0493 & \textbf{0.9205±0.0326} & 89.20±10.73 & 83.92±12.48 & 5.45±5.41 & 0.8814±0.0650 \\
   BL+JCAM & 0.8584±0.0448 & 0.9027±0.0388 & 88.71±7.83 & 82.93±10.90 & 5.33±4.09 & 0.8710±0.0554 \\
   BL+Fuzzy+JCAM & 0.8625±0.0414 & 0.9039±0.0366 & \textbf{93.95±7.96} & \textbf{90.64±10.42} & \textbf{4.92±4.22} & 0.8913±0.0527 \\
   BL+Fuzzy+$\rm CCF_{s}$ & 0.8701±0.0522 & 0.9112±0.0315 & 89.63±3.27 & 85.24±12.53 & 5.71±5.82 & 0.8816±0.0412\\
   BL+Fuzzy+JCAM+$\rm CCF_{s}$ & 0.8738±0.0445 & 0.9187±0.0320 & 92.71±7.93 & 89.01±10.3 & 5.22±4.50 & \textbf{0.8969±0.0554} \\
   \bottomrule
\end{tabular}
\begin{tablenotes}
\footnotesize
\item \% indicates the ratio metrics, \textbf{DLR}: detected length ratio, \textbf{DBR}: detected branch ratio, \textbf{AMR}: airway missing ratio. Model abbreviations \textbf{BL}: baseline 3D UNet with deep supervision, \textbf{JCAM}: jaccard continuity and accumulation mapping loss, \textbf{$\rm CCF_{s}$}: continuity and completeness f-score.
\end{tablenotes}
\end{threeparttable}
\vspace{-1.5em}
\end{table*}
It can be observed that the baseline 3D U-Net could achieve competitive airway segmentation performance, with a 0.8749 IoU score, 87.81\% DLR and 81.74\% DBR. By integrating the fuzzy attention layer with 3D-UNet, a considerable improvement the DBR score was witnessed (roughly 2.2\% average gain). Introducing the JCAM loss promotes the DBR ratio with 82.93\%, while adopting the CCF-score also helps to find a better module (with a 1.3\% increment of DBR). The largest improvement is witnessed by adopting both fuzzy attention layer and JCAM loss, with 90.64\% DBR and 92.95\% DLR and the lowest AMR of 4.92\%. By integrating all these novel strategies (BL+Fuzzy+JCAM+$\rm CCF_s$),  the proposed FANN achieves the highest CCF-score 0.8969, and competitive DBR (89.01\%), DLR (92.17\%) as well as the AMR (5.22\%), while a reduction of IoU and dice score is observed.\\
\textit{\textbf{Comparison experiments on BAS dataset}}. We note that the proposed FANN achieves the state-of-the-art performance for airway segmentation on BAS (Table III), with a 0.8969 CCF-score, 92.71\% DLR, and 89.01\% DBR.  
\begin{table*}[htb]
\centering
\caption{Comparison experiments on the open dataset BAS dataset}
\label{tab:3}  
\begin{threeparttable}
\begin{tabular}{lllllll}
   \toprule
   Model & IoU & Precision & DLR(\%) & DBR(\%) & AMR(\%) & $\rm CCF_{s}$ \\
   \midrule
   Attention UNet\cite{26-oktay2018attention}$^\dagger$ & 0.8762±0.0414 & 0.9207±0.0309 & 88.61±8.32$^\ddagger$ & 82.43±10.65$^\ddagger$ & 5.16±4.18 & 0.8806±0.0534$^\ddagger$ \\
   Juarez et al.\cite{30-garcia2018automatic}$^\dagger$ & 0.8371±0.0752$^\ddagger$ & 0.9344±0.0314 & 73.46±14.19$^\ddagger$ & 64.25±15.55$^\ddagger$ & 11.17±7.62$^\ddagger$ & 0.7879±0.0952$^\ddagger$ \\
   nnUNet\cite{28-isensee2021nnu}$^*$ & \textbf{0.8805±0.0313} & 0.9436±0.0234$^\ddagger$ & 86.84±7.00$^\ddagger$ & 79.21±9.43$^\ddagger$ & 6.96±4.02$^\ddagger$ & 0.8750±0.0416$^\ddagger$ \\
   V-Net\cite{29-milletari2016v}$^*$ & 0.7140±0.1716$^\ddagger$ & \textbf{0.9781±0.0150$^\ddagger$} & 33.96±17.96$^\ddagger$ & 22.04±14.22$^\ddagger$ & 27.15±17.99$^\ddagger$ & 0.4779±0.1751$^\ddagger$\\
   Wang et al.\cite{31-wang2019tubular}$^\dagger$ & 0.7330±0.0786$^\ddagger$ &  0.7636±0.0737$^\ddagger$ & 85.05±12.27$^\ddagger$ & 78.58±14.20$^\ddagger$ & 5.07±6.19 & 0.7813±0.0937$^\ddagger$ \\
   WingNet\cite{16-zheng2021alleviating}$^*$ & 0.8445±0.0596$^\ddagger$ & 0.9373±0.0303$^\ddagger$ & 88.00±11.95‡$^\ddagger$ & 83.11±13.46$^\ddagger$  & 10.32±6.89$^\ddagger$ & 0.8600±0.0768$^\ddagger$ \\
   Proposed & 0.8738±0.0445 & 0.9187±0.0320 & \textbf{92.71±7.93} & \textbf{89.01±10.3} & 5.22±4.50 & \textbf{0.8969±0.0554} \\
   \bottomrule
\end{tabular}
\begin{tablenotes}
\footnotesize
\item * refers to results obtained by an open-source module (with trained weights) or modules trained from open-source codes. † indicates the reproduced results. ‡ represents statistical significance (with Wilcoxon signed-rank test p-value $\textless$ 0.05) compared with the proposed method.
\end{tablenotes}
\end{threeparttable}
\vspace{-1.5em}
\end{table*}
Both the V-Net \cite{29-milletari2016v} and nnUNet \cite{28-isensee2021nnu} achieve high voxel-wise precision (0.9781 and 0.9436, respectively), while the V-Net represent server discontinuity of airway predictions, with a high AMR (27.15\%), low DLR (33.96\%) and DBR (22.04\%). WingNet \cite{16-zheng2021alleviating} achieves a competitive results (88.00\% DLR, 83.11\% DBR and 0.8445 IoU), followed by the Attention UNet \cite{26-oktay2018attention} (88.61\% DLR, 82.43\% DBR and 0.8762 IoU) and the study proposed by Wang et al. \cite{31-wang2019tubular} (85.05\% DLR, 78.58\% DBR and 0.7330 IoU). The study proposed by Juarez et al. \cite{11-garcia2018automatic} achieves good results on IoU (0.8371) and precision (0.9344) but lower DLR (73.46\%) and DBR (64.25\%).\\
\textit{\textbf{Comparison experiments on COVID-19 data}}. The proposed method achieved the best CCF-score (0.9270) among all the comparisons on COVID-19 data, with 93.30\% DLR, 90.18\% DBR, and 2.37\% AMR (shown in Table IV). The nnUNet  \cite{28-isensee2021nnu} achieves the second highest CCF-score of 0.9148, with 91.36\% DLR and 87.33\% DBR. The Attention UNet \cite{26-oktay2018attention} also presents comparable results, with the highest IoU of 0.923, DLR of 88.53\% and DBR of 82.97\%. WingNet \cite{16-zheng2021alleviating} achieves the similar performance of DLR (92.50\%) and DBR (90.67\%) compared with the proposed FAAN, while its IoU (0.8282) is significantly lower than FANN. Unexpectedly, V-Net \cite{29-milletari2016v} still performs poorly on the airway segmentation task, with 34.96\% DLR and 24.07\% DBR\\
\textit{\textbf{Comparison experiments on fibrosis data}}. The performance of all the models witnesses a dramatic reduction in data acquired from patients with fibrotic lung disease (Table IV). The proposed FANN represents superior performance and generalization ability, with a 0.8099 CCF-score, 78.98\% DLR, 73.44\% DBR, and 0.8904 precision. The attention UNet \cite{26-oktay2018attention} obtained a good performance with 72.83 DLR, 66.88\% DBR, and 0.8844 precision, followed by Wing Net \cite{16-zheng2021alleviating} with 69.46\% DLR, 63.01\% DBR, 0.9401 precision and 21.98\% AMR. The nnUNet \cite{28-isensee2021nnu} is failed to obtain comparable results on fibrosis data, with a low DLR (58.15\%) and DBR (50.18\%), indicating a poor continuity of airway prediction of the branches. Although the study proposed by Wang et al. \cite{31-wang2019tubular} achieves a similar DBR and DLR compared with WingNet, their predictions are prone to airway leakages (a low precision of 0.7468).\\ 
\vspace{-0.3cm} 
\begin{table*}[htb]
\centering
\caption{Comparison experiments on our in-house dataset}
\label{tab:4}  
\begin{threeparttable}
\begin{tabular}{clllllll}
   \toprule
   ~ & Model & IoU & Precision & DLR(\%) & DBR(\%) & AMR(\%) & $\rm CCF_{s}$ \\
   \midrule
   \multirow{7}*{COVID-19} 
   ~ & Attention UNet\cite{26-oktay2018attention}$^\dagger$ & \textbf{0.9230±0.0279} & 0.9464±0.0175 & 88.53±6.66$^\ddagger$ & 82.97±8.34$^\ddagger$ & 2.62±2.21 & 0.9057±0.0377$^\ddagger$ \\
   ~ & Juarez et al.\cite{30-garcia2018automatic}$^\dagger$ & 0.8637±0.1016$^\ddagger$ & 0.9566±0.0134 & 71.73±15.07$^\ddagger$ & 63.02±15.07$^\ddagger$ & 10.08±10.79$^\ddagger$ & 0.7914±0.1189$^\ddagger$ \\
   ~ & nnUNet\cite{28-isensee2021nnu}$^*$ & 0.9158±0.0381$^\ddagger$ & 0.9685±0.0407$^\ddagger$ & 91.36±6.72$^\ddagger$ & 87.33±8.89$^\ddagger$ & 2.61±1.73$^\ddagger$ & 0.9148±0.0473$^\ddagger$ \\
   ~ & V-Net\cite{29-milletari2016v}$^*$ & 0.7898±0.0871$^\ddagger$ & \textbf{0.9893±0.0040$^\ddagger$} & 34.96±9.66$^\ddagger$ & 24.07±8.80$^\ddagger$ & 20.33±8.86$^\ddagger$ & 0.5052±0.0911$^\ddagger$\\
   ~ & Wang et al.\cite{31-wang2019tubular}$^\dagger$ & 0.7433±0.1010$^\ddagger$ &  0.7749±0.0736$^\ddagger$ & 84.87±13.20$^\ddagger$ & 79.93±14.09$^\ddagger$ & 5.41±10.35$^\ddagger$ & 0.7870±0.1129$^\ddagger$ \\
   ~ & WingNet\cite{16-zheng2021alleviating}$^*$ & 0.8282±0.0672$^\ddagger$ & 0.9537±0.0215$^\ddagger$ & 92.50±6.89 & \textbf{90.67±8.95} & 13.68±7.15$^\ddagger$ & 0.8689±0.0680$^\ddagger$ \\
   ~ & \textbf{Proposed} & 0.9222±0.0261 & 0.9431±0.0166 & \textbf{93.30±5.29} & 90.18±7.59 & \textbf{2.37±1.64} & \textbf{0.9270±0.0338} \\
   \cmidrule(lr){1-8}
   \multirow{7}*{Fibrosis} 
   ~ & Attention UNet\cite{26-oktay2018attention}$^\dagger$ & 0.8202±0.0553 & 0.8844±0.0640 & 72.83±9.41$^\ddagger$ & 66.88±10.84$^\ddagger$ & 7.98±2.97 & 0.7764±0.0678$^\ddagger$ \\
   ~ & Juarez et al.\cite{30-garcia2018automatic}$^\dagger$ & 0.7911±0.0532$^\ddagger$ & 0.9296±0.0294$^\ddagger$ & 56.16±10.76$^\ddagger$ & 48.29±12.18$^\ddagger$ & 15.68±6.37$^\ddagger$ & 0.6688±0.0688$^\ddagger$ \\
   ~ & nnUNet\cite{28-isensee2021nnu}$^*$ & \textbf{0.8312±0.0495}$^\ddagger$ & 0.9381±0.0314$^\ddagger$ & 58.15±6.80$^\ddagger$ & 50.18±7.93$^\ddagger$ & 11.74±2.93$^\ddagger$ & 0.6972±0.0564$^\ddagger$ \\
   ~ & V-Net\cite{29-milletari2016v}$^*$ & 0.5002±0.0967$^\ddagger$ & \textbf{0.9768±0.0006}$^\ddagger$ & 8.54±3.76$^\ddagger$ & 3.40±2.99$^\ddagger$ & 49.36±9.88$^\ddagger$ & 0.1576±0.0568$^\ddagger$\\
   ~ & Wang et al.\cite{31-wang2019tubular}$^\dagger$ & 0.6979±0.0647$^\ddagger$ &  0.7468±0.0773$^\ddagger$ & 69.61±9.24$^\ddagger$ & 62.61±11.17$^\ddagger$ & 8.22±3.88 & 0.6971±0.0747$^\ddagger$ \\
   ~ & WingNet\cite{16-zheng2021alleviating}$^*$ & 0.7436±0.0805$^\ddagger$ & 0.9401±0.0156$^\ddagger$ & 69.46±9.71$^\ddagger$ & 63.01±11.39$^\ddagger$  & 21.98±8.48$^\ddagger$ & 0.7208±0.0872$^\ddagger$ \\
   ~ & \textbf{Proposed} & 0.8269±0.0402 & 0.8904±0.0373 & \textbf{78.98±8.00} & \textbf{73.44±9.54} & \textbf{7.95±2.37} & \textbf{0.8099±0.0517} \\
   
   \bottomrule
\end{tabular}
\begin{tablenotes}
\footnotesize
\item * refers to results obtained by open-source module (with trained weights) or module trained from open-source codes. † indicates the reproduced results. ‡ represents statistical significance (with Wilcoxon signed-rank test p-value $\textless$ 0.05) compared with the proposed method.
\end{tablenotes}
\end{threeparttable}
\vspace{-1.5em}
\end{table*}
\section{Discussion}
This section mainly discusses the overall performance, the importance of data sampling, the airway leakage and neglect in the prediction, the importance of evaluation metrics, and the detection rate of different sized branches.
\noindent \textbf{Overall performance analysis.} The experimental results on public and our in-house datasets have demonstrated the superior performance and robustness of the proposed method for airway segmentation. Among all the comparisons, V-Net obtained the poorest performance, due to the discontinuity of airway predictions (all the predictions were post-processed by keeping the largest 3D component). Here, we mainly discuss three comparison models with the proposed method, including attention UNet, nnUNet, and WingNet. In all three sub-datasets, both nnUNet and WingNet represented competitive precision compared to FANN and attention UNet. However, the AMR score of WingNet was roughly twice larger than that of the nnUNet, while the DBR and DLR of the WingNet were better than that of the nnUNet. This illustrates that WingNet obtained worse predictions on the main trachea, but better predictions on small/middle-sized branches. Compared with the other two models, the attention UNet presents better performance on all the datasets, showing the effectiveness of the attention mechanism. \\
\indent In particular, FANN achieved significantly better results in DLR, DBR, and CCF-scores than nnUnet, attention UNet and WingNet in the public BAS dataset. The IoU of FANN (0.8738) represented no significant differences compared with nnUNet (0.8805) and attention UNet (0.8762) which was much better than that of the WingNet (0.8445). In the COVID data, the proposed FANN gained better DLR and DBR compared to nnUNet and attention UNet. Although WingNet achieves a relatively low IoU (0.8282), it obtains a similar DLR and DBR to FANN (with Wilcoxon signed-rank test p-value \textgreater 0.05). On fibrosis data, although the nnUNet achieved a better IoU and precision score than FANN, it suffers from heavy discontinuity with poor DLR (58.15\%) and DBR (50.18\%) gained. WingNet achieves the highest precision (0.9401) but the largest AMR (21.98\%), indicating severe false-negative predictions. FANN achieves the best CCF-score, DLR, DBR, and AMR among all the comparisons, with competitive IoU and precision scores.  \\
\textbf{Data sampling strategy.} The strategy of data sampling is prone to be overlooked for patch-based modules. In this paper, we presented a smart data sampling strategy for patch extraction in the imbalanced dataset. To better explore the importance of data sampling, we conducted experiments on the open BAS dataset in terms of different data sampling strategies, including different extraction criteria and patch sizes. Here we define three extraction criteria, including (1) Sequential: extracting the patches sequentially by sliding windows within the minimum bounding box volume of the 3D ground truth area; (2) Drop: extracting the patches sequentially by sliding windows within the minimum bounding box volume of the 3D ground truth area and discarding the main trachea (Drop-A) or pure negative patches (Drop-B) to alleviate data imbalance. (3) Smart patch sampling as described in section 3.3. We also compare the patch size defined by three different criteria, with (1) uniform of 128×128×128 (2) the roughly same ratio as the median size of the ground truth volumes (160×96×160) (3) the roughly same ratio as the mean size of the ground truth volumes (128×96×144). 
\begin{table*}[ht]
\centering
\caption{Comparison experiments of data sampling strategy}
\label{tab:5}  
\begin{threeparttable}
\begin{tabular}{llllllll}
   \toprule
   ~ & Extraction  & Patch size & IoU & Precision & DLR(\%) & DBR(\%) & AMR(\%) \\
   \midrule
   Cut-1 & Sequential & uniform & 0.8782±0.0641 & 0.9341±0.0409 & 85.17±14.28 & 76.41±15.27 & 7.64±7.22 \\
   Cut-2 & Sequential & median & 0.8854±0.0470 & 0.9408±0.0300 & 85.79±8.12 & 78.77±10.06 & 6.03±4.63 \\
   Cut-3 & Sequential & mean & 0.8862±0.0422 & 0.9394±0.0275 & 86.66±8.88 & 79.39±10.80 & 6.06±4.61 \\
   Cut-4 & Drop-A & mean & 0.8069±0.0826 & 0.9437±0.0316 & 84.64±10.00 & 77.57±12.70 & 14.87±8.13\\
   Cut-5 & Drop-B & mean & \textbf{0.8883±0.0407} & \textbf{0.9451±0.0286} & 87.73±8.01 & 80.92±10.51 & 5.95±4.25 \\
   Cut-6 & Smart & mean & 0.8749±0.0385 & 0.9190±0.0302 & \textbf{87.81±8.74} & \textbf{81.74±11.15} & \textbf{5.11±4.10} \\
   \bottomrule
\end{tabular}
\begin{tablenotes}
\footnotesize
\item The ‘uniform’ patch size refers to 128×128×128, while the ‘median’ and ‘mean’ are 160×96×160 and 128×96×144, respectively.
\end{tablenotes}
\end{threeparttable}
\vspace{-1.5em}
\end{table*}
\indent Although little increment of IoU is emerged from Cut-1, Cut-2, and Cut-3, a significant improvement in the DLR and DBR is noted. In addition, the results of Cut-3 and Cut-4 (Table V) indicate that the segmentation performance can be improved by discarding the pure negative patches (with 1.6\% and 1\% gain in DBR and DLR, respectively). However, discarding the main trachea may not be applicable since the module failed to predict the trachea by learning the features of small and medium-size branches (shown in Cut-4). The model trained on Cut-5 presents a competitive DBR with a roughly 4.5\% increment compared with Cut-1. The proposed SPS strategy achieves the best performance with the highest DLR, DBR and lowest AMR. \\
\begin{figure*}[hb]
\vspace{-0.3cm} 
\includegraphics[scale=0.45]{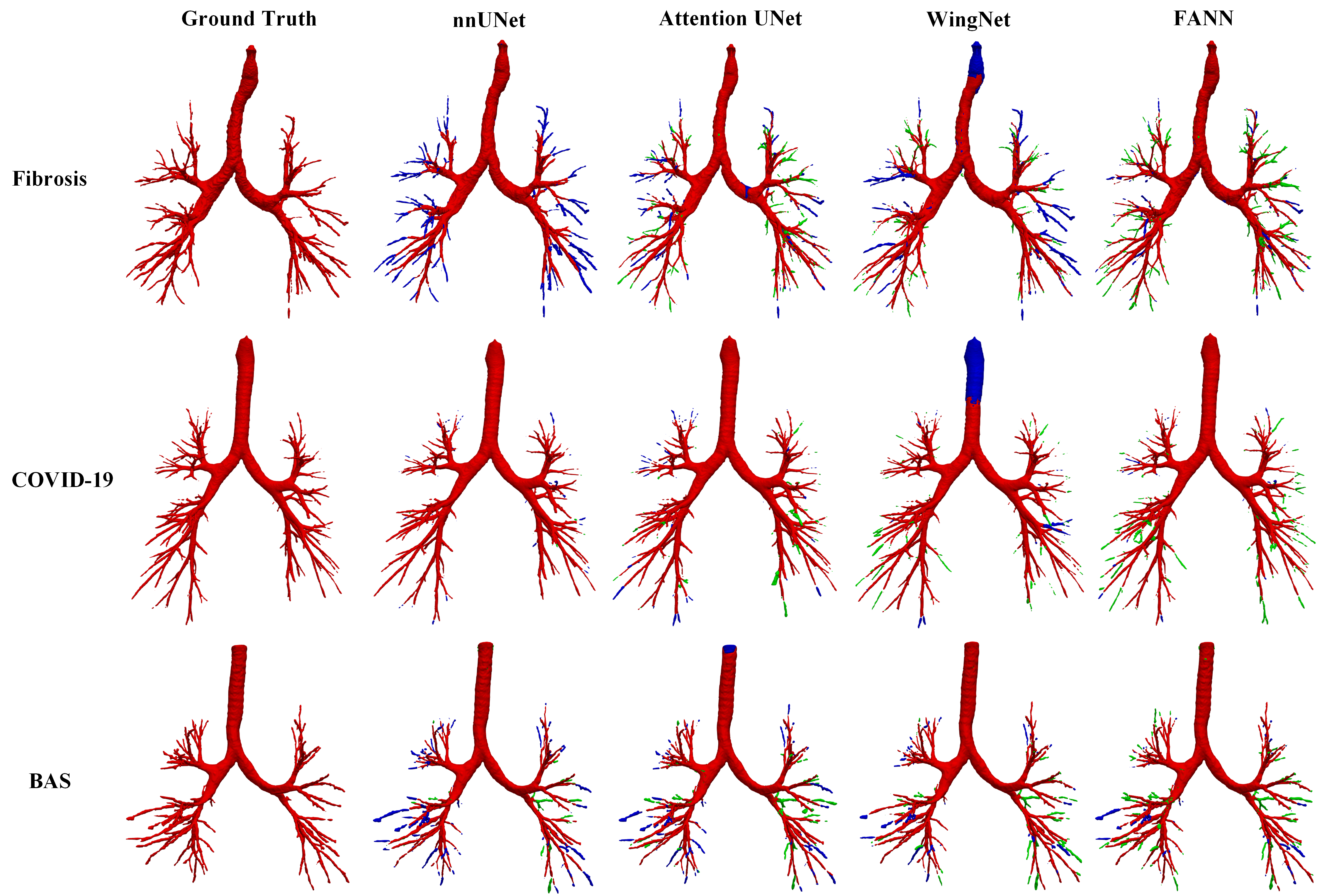}      
\centering
\caption{Visualization results of the competitive module, including nnUNet, Attention UNet, WingNet and the proposed FANN on fibrosis data (first row), COVID-19 data (second row), and BAS data (third row), with the red, blue, green color representing the true positive, false negative (airway neglects), and false positive (airway leakages) predictions, respectively.}
\vspace{-0.3cm} 
\label{fig5}
\end{figure*}
\noindent\textbf{Airway leakages and Airway neglect. } Airway leakages are false-positive predictions that can be reflected by the precision score. Airway neglect refers to false-negative predictions that are assessed by AMR. As shown in Fig. 5, due to the unavoidable neglect of manual labelling, some leakages might be ascribed to the airway branches that are not annotated correctly. nnUNet presents a high precision score with limited airway leakages, while it suffers from relative AMR with many airway neglects (especially in fibrosis cases). WingNet performs severe airway neglect of the main trachea, leading to a low IoU score with comparable DBR and DLR. WingNet, attention UNet, and FANN present a few leakages at the terminal branches, indicating the incompleteness of airway annotations. Compared with all the studied models, FANN achieves the smallest AMR and most of its leakages belong to incomplete annotations. Therefore, we reach the conclusion that a small amount of airway leakages are encouraged, as they can help the model better predict terminal bronchioles.  \\
\textbf{Evaluation metric.} The conventional evaluation metrics such as IoU score or dice score no longer work well to find the best module during the training. As Tables II, III and IV illustrated, the module with the highest IoU score does not refer to the one with the best segmentation ability. For instance, the nnUNet achieved the best IoU score on the BAS dataset, but it obtained a weak performance on the fibrosis evaluation. Meanwhile, only focusing on the branch performance while ignoring the overlapped metric may also lead to airway leakages. For example, the BL+fuzzy+JCAM achieved better DLR and DBR but a lower CCF-score than the final combination in the ablation studies of FAAN (Table I). To explore the most appropriate metric, we further evaluated the BL+fuzzy+JCAM on the fibrosis and COVID-19 datasets. It achieved 0.7961(0.9099) IoU score, 74.66\% (93.81\%) DLR and 68.85\% (91.87\%) DBR, 10.29\% (2.36\%) AMR, and 0.7687 (0.9215) CCF-score on fibrosis (COVID-19) data, with a considerable gap compared with the FAAN. This also proves the efficiency of the proposed CCF-score, which can better help the module selection.\\
\begin{table}[ht]
\setlength{\belowcaptionskip}{-1.5cm}
\centering
\caption{Detection rate of different-sized branches}
\label{tab:6}  
\begin{threeparttable}
\begin{tabular}{lllll}
   \toprule
   ~ & TB  & SB & MB & LB \\
   \midrule
   BAS & 81.55\% & 95.76\% &98.59\%	&99.38\% \\
   COVID-19 & 82.96\% & 97.05\% & 99.67\% & 100\% \\
   Fibrosis	& 47.15\% & 87.60\% & 98.32\% & 98.50\% \\
   \bottomrule
\end{tabular}
\end{threeparttable}
\vspace{-2em}
\end{table}
\textbf{The detection rate of different-sized branches.} To explore the segmentation performance of different-sized airway branches, the detection rate of the branches with different radius is calculated. Here we define a certain branch is correctly detected when its IoU is larger than 0.8. We simply divided the airway branches into 4 sizes based on their estimated radius, including TB (terminal branches that range from 0 to 2mm), SB (small branches 2mm to 4mm), MB (middle branches range from 4mm to 8mm) and LB (large branches that larger than 8mm). The performance of the proposed FAAN on BAS, COVID-19 and fibrosis datasets is shown in Table VI. Table V illustrates that the proposed method has achieved superior performance of airway segmentation on BAS and COVID-19 data, while the terminal branch detection rate on the Fibrosis data is relatively low. This is caused by the complex honeycombing in fibrotic lung disease which has similar structural similarity with the airway branches. \\
\textbf{Potential studies.} This study has shown the feasibility of adopting fuzzy logic into deep attention neural networks. In particular, many potential research directions can be explored. E.g., introducing a fuzzy cost function for better optimization, or bringing in other MFs such as generalized bell and sigmoid MFs.

\section{Conclusion}
\noindent Airway segmentation is a crucial step that can help clinicians better assess disease and prognosis analysis. However, most existing methods heavily suffer from discontinuous prediction for small-sized airway branches. The proposed study illustrates the importance of data sampling strategy and evaluation metrics. The effectiveness of fuzzy attention neural network, Jaccard continuity and accumulation mapping loss, and the CCF-score for 3D airway segmentation has been demonstrated. The superior performance of the proposed method has been observed on both open datasets and our in-house datasets, including scans acquired from patients with cancer, COVID-19, fibrosis, and mild lung disease. 

The imperfect terminal branches detection rate on fibrosis data indicates the potential studies. In the future, we will conduct further studies that integrate fuzzing learning and deep learning to improve the detection rate of tiny airway branches in patients with fibrotic lung disease.
 
\section{Acknowledgement}
This study was supported in part by the Basque Government (KK-2020/00049), Consolidated Research Group MATHMODE (IT1294-19), the National Natural Science Foundation of China (62076182), the BHF (TG/18/5/34111, PG/16/78/32402), the ERC IMI (101005122), the H2020 (952172), the MRC (MC/PC/21013), the SABER project supported by Boehringer Ingelheim Ltd, and the UKRI Future Leaders Fellowship (MR/V023799/1). We thank Prof Jun Xia from the Department of Radiology, Shenzhen Second People’s Hospital providing part of the data with ethics approval incorporated in this study.

\bibliography{ref}
\end{document}